\documentclass[10pt]{article}
\usepackage{cite}
\usepackage{graphicx}
\usepackage{url}
\usepackage{amsmath,amssymb,amsfonts}
\usepackage{subfigure}
\usepackage{wrapfig}
\usepackage{fancyhdr}
\usepackage{authblk}
\renewcommand{\thispagestyle}[1]{} 

\author[1]{V. Kulathumani\thanks{vinod.kulathumani@mail.wvu.edu}}
\author[1]{M. Nakagawa\thanks{manakagawa@mix.wvu.edu}}
\author[2]{A. Arora\thanks{anish.arora@samraksh.com}}
\affil[1]{Department of Computer Science, West Virginia University, Morgantown WV 26505}
\affil[2]{The Samraksh Company, Dublin OH 43017}

\begin{document}
\title{Coverage characteristics of self-repelling random walks in mobile ad-hoc networks}

\maketitle

\begin{abstract}
A self-repelling random walk of a token on a graph is one in which at each step, the token moves to a neighbor that has been visited least often (with ties broken randomly). The properties of self-repelling random walks have been analyzed for two dimensional lattices and these walks have been shown to exhibit a remarkable uniformity with which they visit nodes in a graph. In this paper, we extend this analysis to self-repelling random walks on mobile networks in which the underlying graph itself is temporally evolving. Using network simulations in ns-3, we characterize the number of times each node is visited from the start until all nodes have been visited at least once. We evaluate under different mobility models and on networks ranging from $100$ to $1000$ nodes. Our results show that until about $85\%$ coverage, duplicate visits are very rare highlighting the efficiency with which a majority of nodes in the network can be visited. Even at $100\%$ coverage, the exploration overhead (the ratio of number of steps to number of unique visited nodes) remains low and under $2$.  Our analysis shows that self-repelling random walks are effective, structure-free tools for data aggregation in mobile ad-hoc networks.
\end{abstract}

\section{Introduction}
A self-repelling random walk is one in which at each step the walk moves towards one of the neighbors that has been least visited \cite{self-repelling}. Self-repelling random walks were introduced in the $1980s$ and have been studied extensively in the physics literature. One of the striking properties of self-repelling random walks is the remarkable uniformity with which they visits nodes in a graph. This has been studied formally in terms of the variance of the number of visits at each node during sufficiently long instances of a self-repelling random walk \cite{rw12}. More precisely, let $n_i(t,x)$ be the number of times a node $i$ has been visited, starting from a node $x$. The quantity studied in \cite{rw12} is the variance $(1/N)(\sum_i (n_i(t,x) - \mu)^2)$, where $\mu = (1/N)(\sum_i  n_i(t,x))$. The study in \cite{rw12} shows that this variance is bounded by values less than $1$ even in lattices of dimensions $2048 \times 2048$. 

In this paper, we seek to extend this analysis for self-repelling random walks that execute on mobile networks (MANETs) where the underlying graph itself is temporally evolving. Our motivation for this study is the applicability of self-repelling random walks for data aggregation problems in MANETs. The goal of data aggregation is to compute statistical summaries (such as max, min, average, count) across nodes in a network. In networks where nodes are static and links are stable, data aggregation can be achieved by collecting data along fixed routing structures such as trees or network backbones \cite{naik_sprinkler, ctp}. However, in mobile networks, routing has proven to be quite challenging beyond scales of a few hundred nodes primarily because topology driven structures are unstable and are likely to incur a high communication overhead for maintenance \cite{rtswjournal}. Therefore, structure-free techniques are more appropriate for data aggregation in mobile networks and random walks are an appropriate candidate for the same. 

The idea in this paper is to introduce a token in the network that successively visits all nodes in the network using a random walk traversal and computes the overall aggregate. Each time a token visits a node, information from the node can be added into the aggregate and the token can be passed to the next node. But, traditional random walks may be too slow in covering all nodes in the network because they may get stuck in regions of already visited nodes. {\em On the other hand, the uniformity property of self-repelling random walks is interesting because it implies that the token is likely to spread towards unvisited areas in the network.} As a result, the aggregate computation can be quite efficient because the number of duplicate visits to the same nodes would be small. Our {\em aim in this paper} is to quantify how efficiently self-repelling random walks traverse nodes in a mobile network starting from zero nodes visited until all nodes have been visited at least once.

It is important to clarify that the tokens used in self-repelling random walks are essentially {\em memory-less}. In other words, the token does not carry information about which nodes have been visited. Instead, the nodes themselves carry this information. Each node locally counts how many times it has been visited. The token following a self-repelling random walk only chooses the "best" node to visit next based on the information it receives at each step. Therefore, {\em the signaling overhead is minimal and does not grow with neighborhood density.} Nodes can suppress their request as soon as they hear another request from a node with fewer and equal number of visits. This is what make self-repelling random walks a very attractive
tool for data aggregation in mobile ad-hoc networks

We simulate self-repelling random walks using ns-3 in mobile ad-hoc networks ranging in size from $100$ to $1000$ nodes under different mobility models and node speeds (described in Section~\ref{sec:model}). We study the coverage characteristics of self-repelling random walks using two metrics. The first is the number of times each node is visited at different phases of the random walk starting from when the walk begins until all nodes have been visited at least once. This metric captures the uniformity with which nodes are visited. The second is the exploration overhead which is defined as the ratio of number of steps of the random walk to the number of unique nodes that are visited in the network. This metric captures the overhead of re-visiting already visited nodes in the network as the token moves towards $100\%$ coverage. Our results show that even at $100\%$ coverage, the exploration overhead remains low highlighting that they can be used for efficient data aggregation in MANETs.

\begin{figure*}[htbp]

  \begin{center}
    \mbox{
      \subfigure[Distribution of number of visits at $50\%$ coverage] {\scalebox{0.4}{\includegraphics[width=\textwidth]{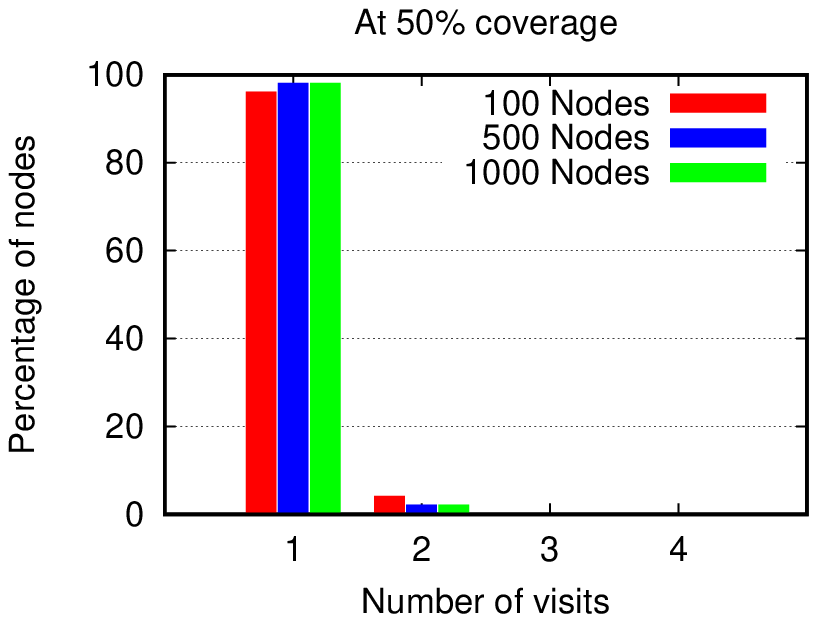}} \label{fig:1a}} \quad
      \subfigure[Distribution of number of visits at $75\%$ coverage] {\scalebox{0.4}{\includegraphics[width=\textwidth]{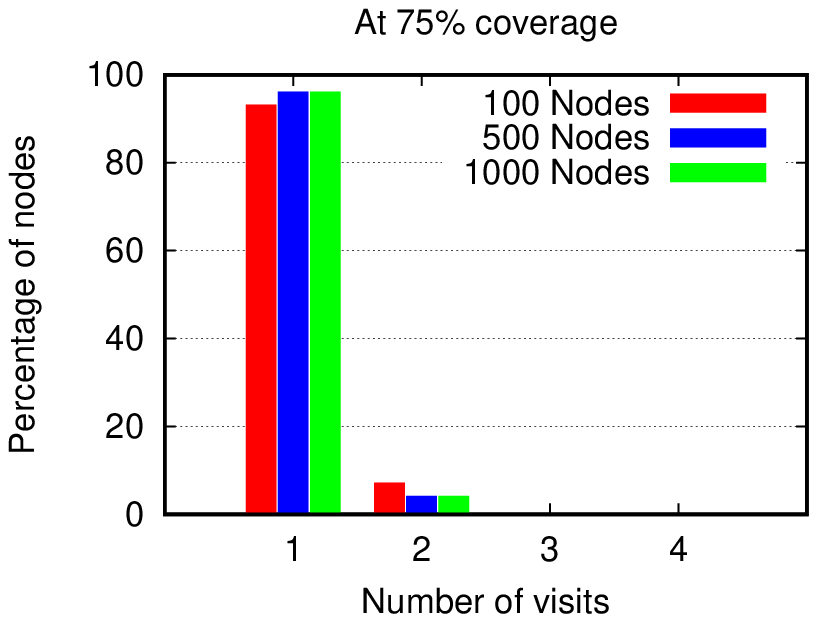}} \label{fig:1b}} 
      } \\
    \mbox{
      \subfigure[Distribution of number of visits at $85\%$ coverage] {\scalebox{0.4}{\includegraphics[width=\textwidth]{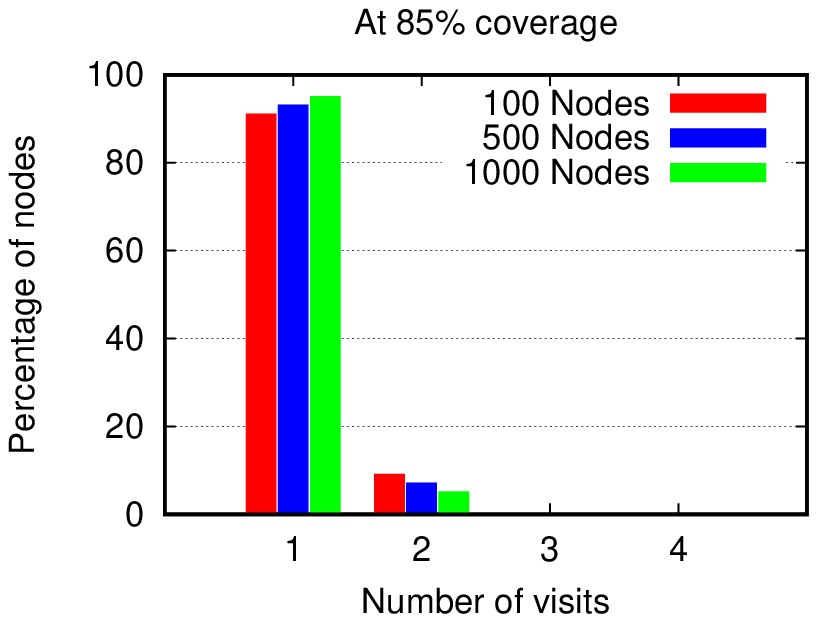}} \label{fig:1c}} \quad
      \subfigure[Distribution of number of visits at $100\%$ coverage] {\scalebox{0.4}{\includegraphics[width=\textwidth]{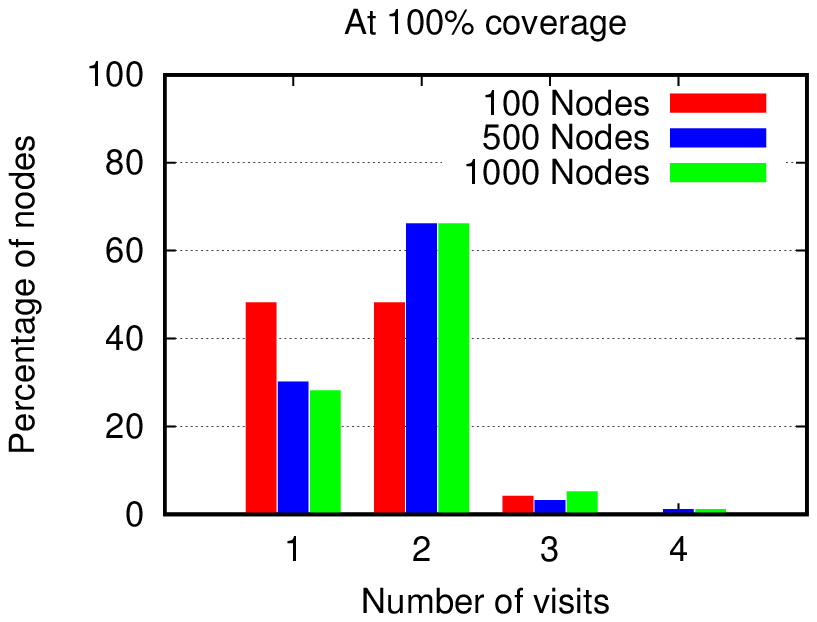}} \label{fig:1d}}     
      } 
    \vspace{-1mm}  
    \caption{Distribution of number of visits at each node at different stages of exploration of a self-repelling random walk (network size 100,400 and 1000 nodes)}
       \label{fig:1}
  \end{center}
\vspace*{-4mm}
\end{figure*}

\section{Model}
\label{sec:model}

We simulate a mobile network of $N$ nodes using ns-3, each with communication range $R$, independently and  uniformly deployed over a square region of sides $\sqrt{A}$ resulting in a network density $\rho = N/A$ of the deployed nodes. We let $R$ to be large enough so that the network remains connected whp. This is referred to as the connectivity threshold in terms of communication range $R$ and has been shown to be $R^2 = \theta(log(N)/\rho)$ \cite{avin-thesis}. For our simulations, we adjust the deployment area and communication range such that $R^2 = log(N)/\rho$. 

We consider $2$ different mobility models for the nodes. The first is a random direction mobility model (with reflection) \cite{rw10} in which, at each interval a node picks a random direction uniformly in the range $[0,2\pi]$ and moves with a constant speed that is randomly chosen in the range $[v_l,v_h]$. At the end of each interval, a new direction and speed are calculated. If the node hits a boundary, the direction is reversed. Motion of the nodes is independent of each other. An important characteristic of this mobility model is that it preserves the uniformity of the distribution of node locations: given that at time $t = 0$ the position and orientation of users are independent and uniform, they remain uniformly distributed for all times $t > 0$ provided the users move independently of each other \cite{rw11}. The other mobility model we consider is 2-d random waypoint where such uniformity assumptions may not hold \cite{rw11}. Here, each mobile node randomly selects one location in the simulation area and then travels towards this destination with constant velocity chosen randomly from $[v_l,v_h]$ \cite{mobilitymodels-survey} . Upon reaching the destination, the node stops for a duration defined by the {\em pause time}. After this duration, it again chooses another random destination and the process is repeated. We set the pause time to $2$ seconds between successive changes. 

We consider average node speeds in the range of $3$ to $15$ m/s. For the deployment density that we have chosen, a mapping between node speed and the average link changes per node per second is listed in Table~\ref{tab1}. This table quantifies the link instability caused by node mobility at different node speeds. As seen in Table~\ref{tab1}, the network structure is rapidly changing at the speeds chosen for evaluation. 

\begin{table}[ht]
\caption{Mapping between node speed and link changes per node per second} 
\centering 
\begin{tabular}{c c c c c} 
\hline\hline 
Size & 3m/s & 7m/s & 11m/s & 15m/s \\ [0.5ex] 
\hline 
100 & 1 & 3 & 5 & 7 \\ 
300 & 2 & 6 & 7 & 9 \\
500 & 3 & 7 & 10 & 12 \\
1000 & 3 & 8 & 12 & 14 \\ [1ex] 
\hline 
\end{tabular}
\label{tab1} 
\end{table}


A token is introduced at a random location in the network and executes a self-repelling random walk. At each step, it moves to a neighbor that has been visited least often (with ties broken uniformly at random). We track the number of times each node has been visited and the exploration overhead (as defined in Section~1) until all nodes have been visited.

\section{Analysis}
\label{sec:analysis}

\begin{figure*}[htbp]

  \begin{center}
    \mbox{
      \subfigure[Exploration overhead as a function of percentage of nodes visits] {\scalebox{0.42}{\includegraphics[width=\textwidth]{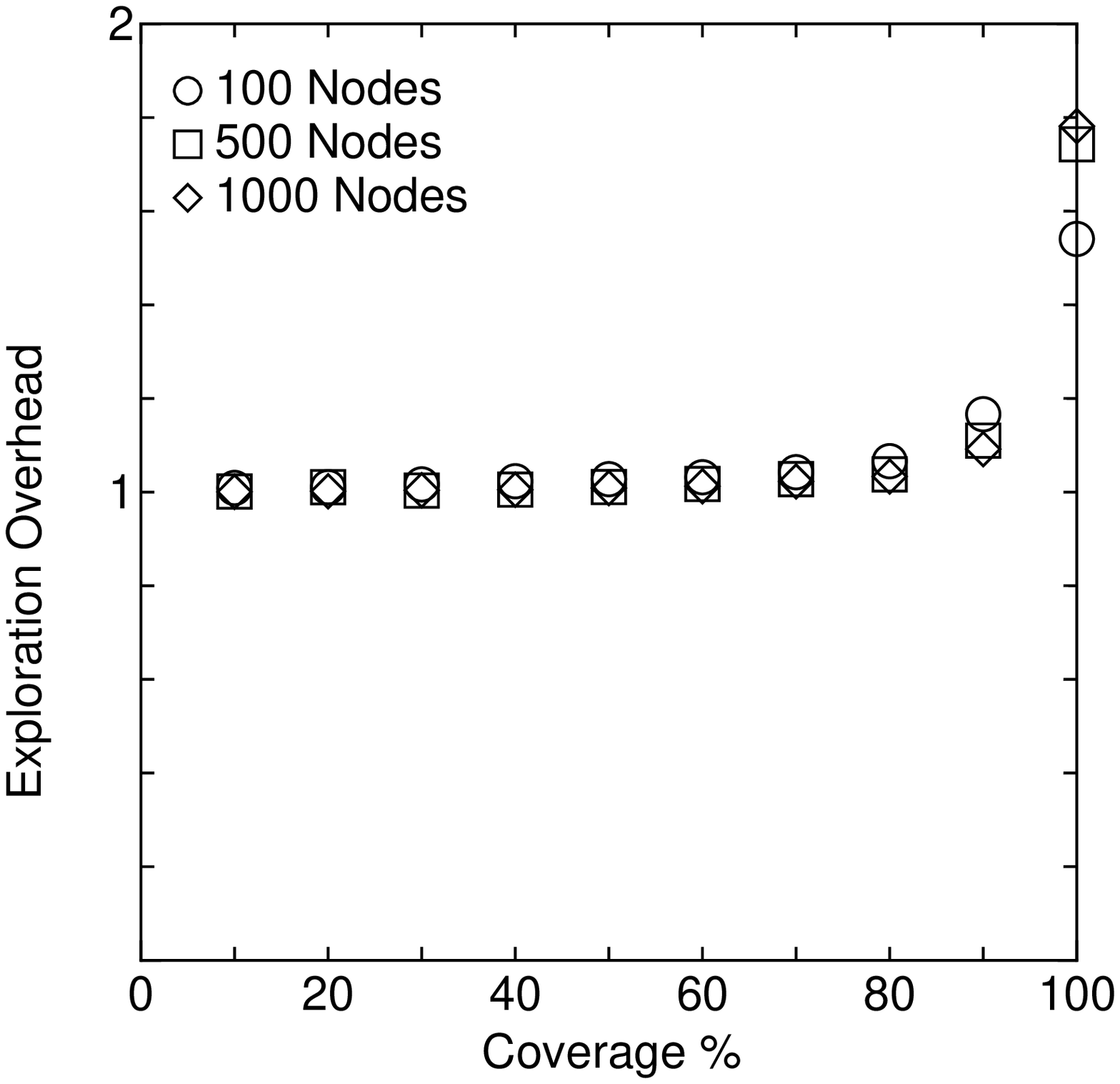}} \label{fig:2a}} \quad
      \subfigure[Exploration overhead at $100\%$ coverage as a function of network size] {\scalebox{0.42}{\includegraphics[width=\textwidth]{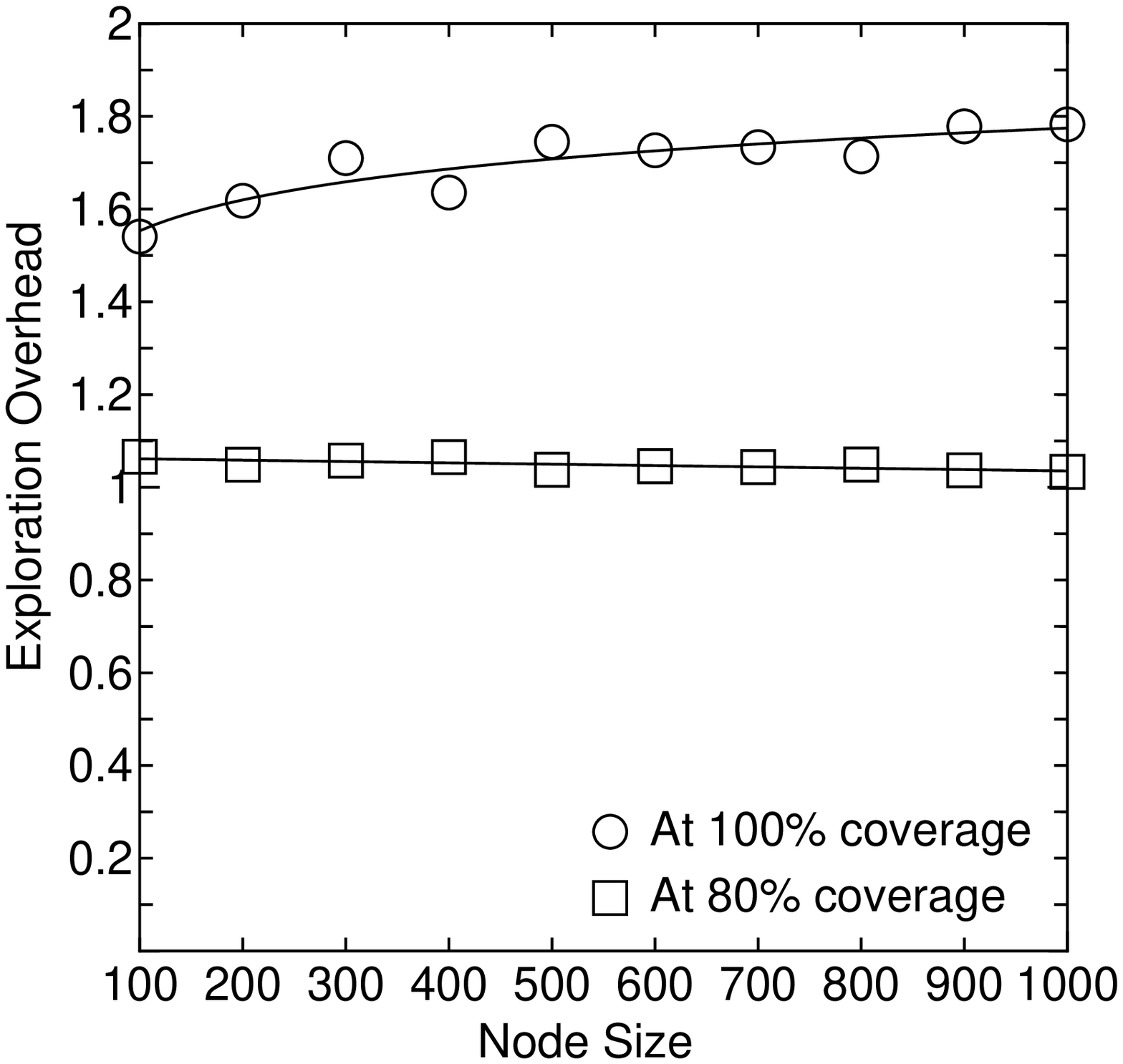}} \label{fig:2b}} 
          
      } 
    \vspace{-1mm}  
    \caption{{Analysis of exploration overhead of a self-repelling random walk in a MANET}}
       \label{fig:2}
  \end{center}
\vspace*{-4mm}
\end{figure*}

\begin{figure*}[htbp]

  \begin{center}
    \mbox{
      \subfigure[Distribution of  number of visits to each node in a pure random walk] {\scalebox{0.42}{\includegraphics[width=\textwidth]{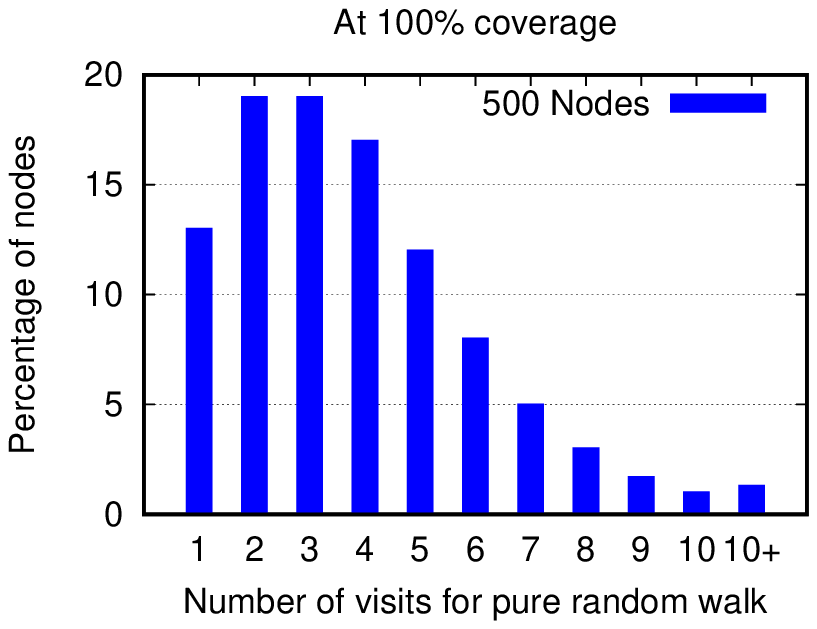}} \label{fig:4a}} \quad
      \subfigure[Distribution of  number of visits to each node in a self-repelling random walk ] {\scalebox{0.42}{\includegraphics[width=\textwidth]{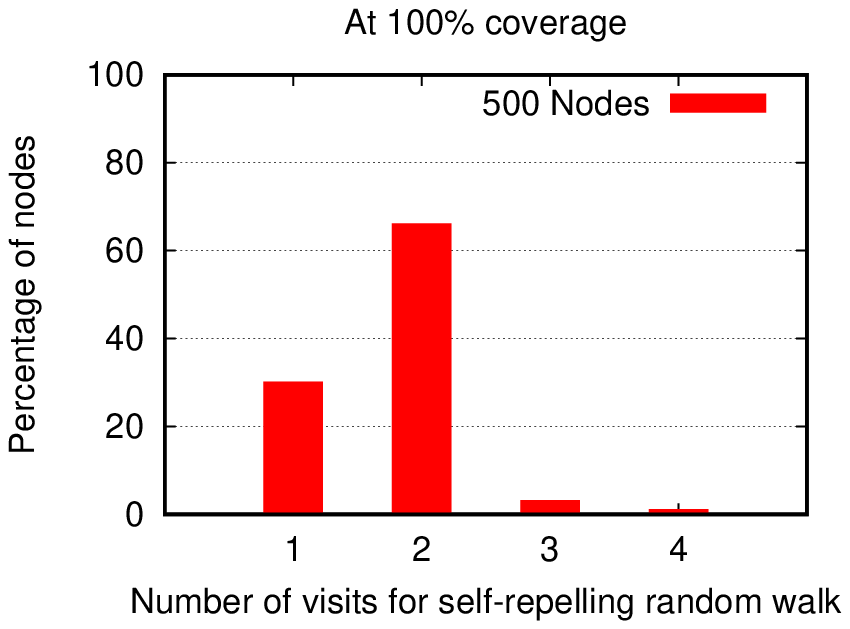}} \label{fig:4b}} 
          
      } 
    \vspace{-1mm}  
    \caption{{Comparison of coverage uniformity with pure random walks}}
       \label{fig:4}
  \end{center}
\end{figure*}

\begin{figure*}[htbp]

  \begin{center}
    \mbox{
      \subfigure[Exploration overhead at full coverage as a function of network size] {\scalebox{0.42}{\includegraphics[width=\textwidth]{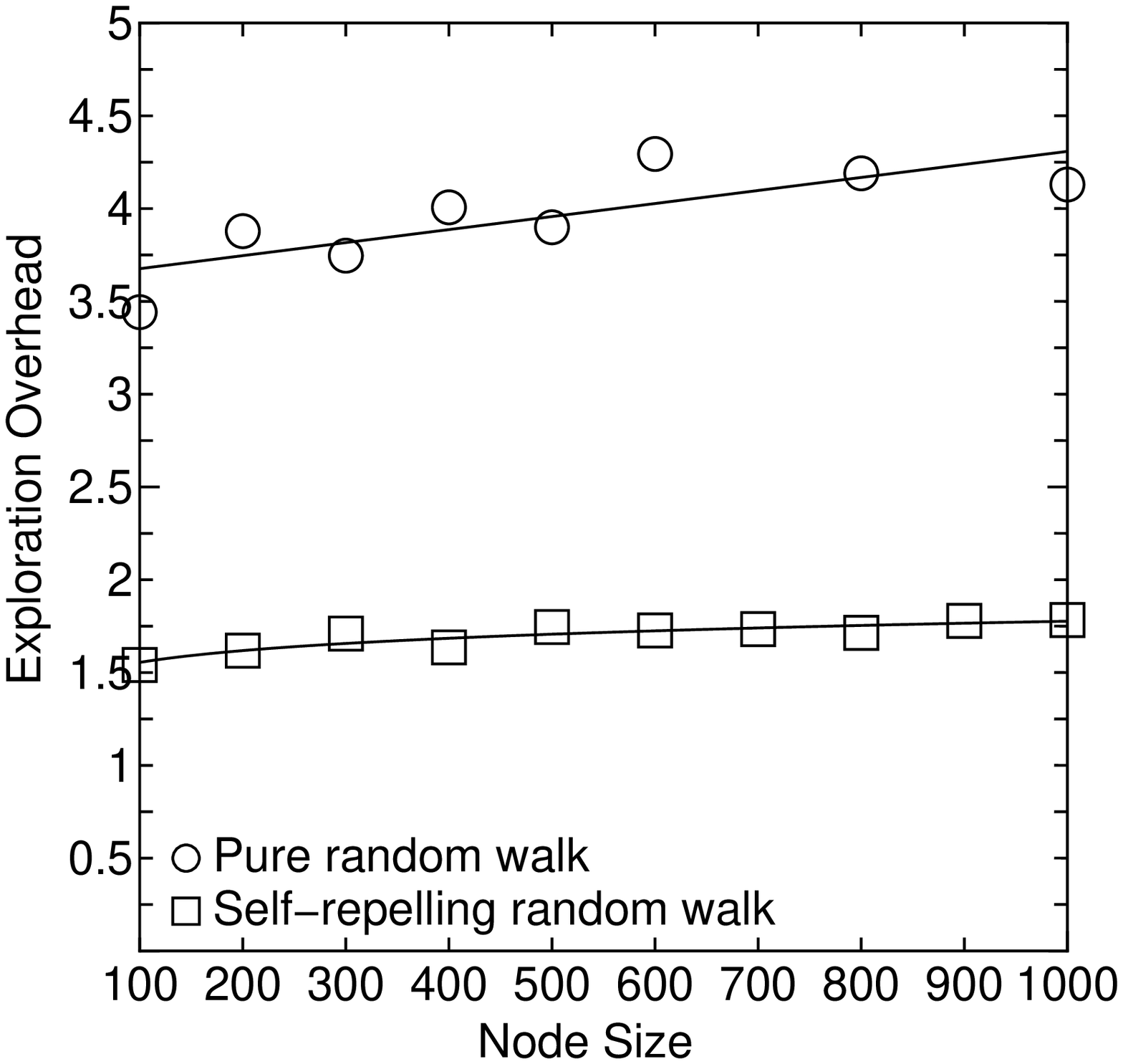}} \label{fig:5a}} \quad
      \subfigure[Exporation overhead as a function of coverage percentage] {\scalebox{0.42}{\includegraphics[width=\textwidth]{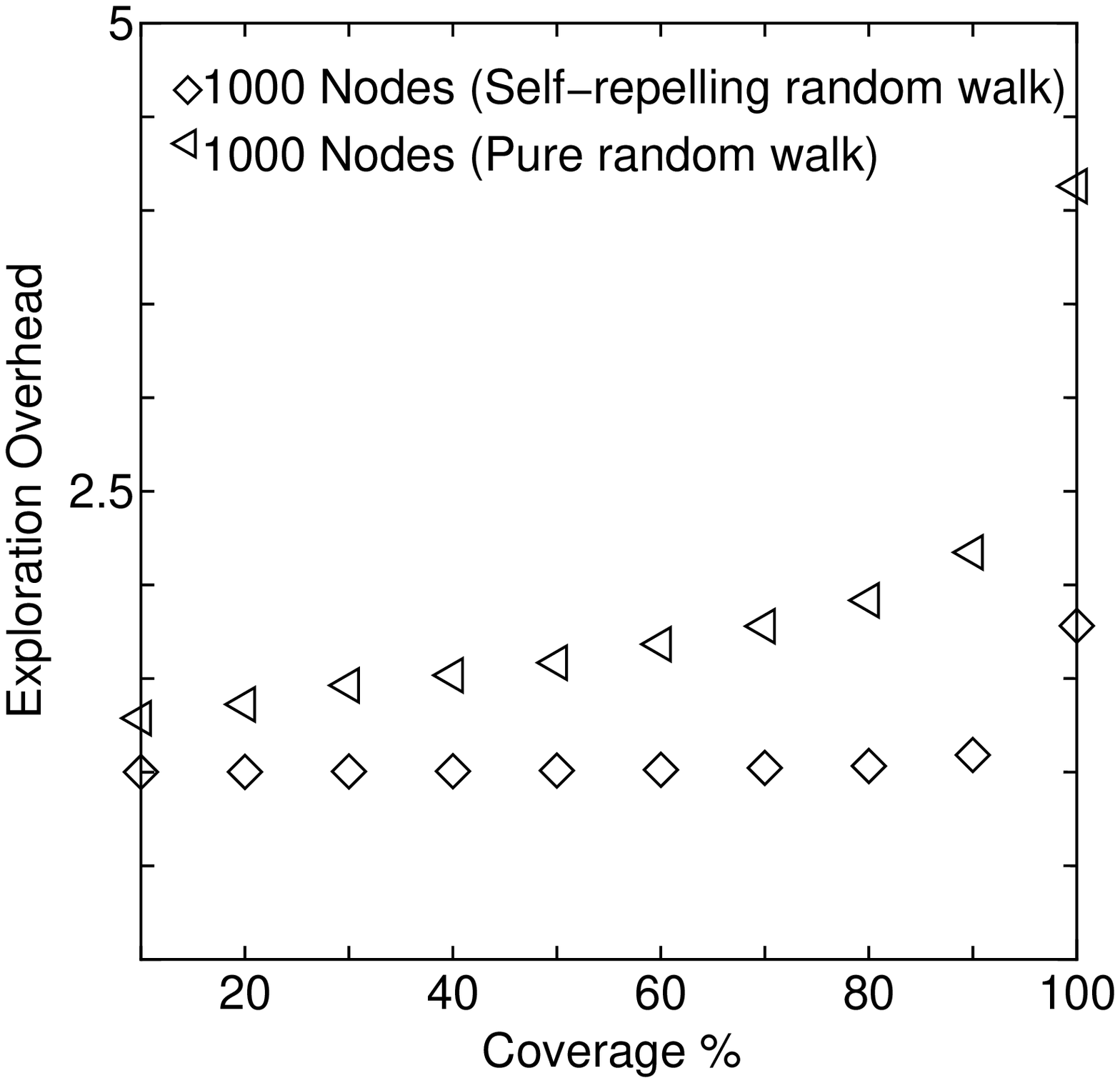}} \label{fig:5b}} 
          
      } 
    \vspace{-1mm}  
    \caption{{Comparison of exploration overhead for pure random walks and self-repelling random walks}}
       \label{fig:5}
  \end{center}
\end{figure*}

\subsection{Coverage uniformity}
In Fig.~\ref{fig:1a}, Fig.~\ref{fig:1b} and Fig.~\ref{fig:1c},we show the number of times each node is visited when the random walk has finished visiting $50\%$ of the nodes, $75\%$ of the nodes and $85\%$ of the nodes. We observe that most of the nodes are just visited once and this result holds even at $1000$ nodes. These graphs highlight the uniformity with which nodes are visited as self-repelling random walks progress. The random walk is not stuck in regions of already visited nodes - instead, it spreads towards unvisited areas. Otherwise, one would have observed more duplicate visits to the previously visited nodes.In Fig.~\ref{fig:1d}, we analyze the distribution of number of visits at each node when $100\%$ coverage is attained. Here, we see that most nodes are visited $2$ or $3$ times and the distribution falls off rapidly after that. 

\begin{figure*}[htbp]

  \begin{center}
    \mbox{
      \subfigure[Impact of mobility models] {\scalebox{0.42}{\includegraphics[width=\textwidth]{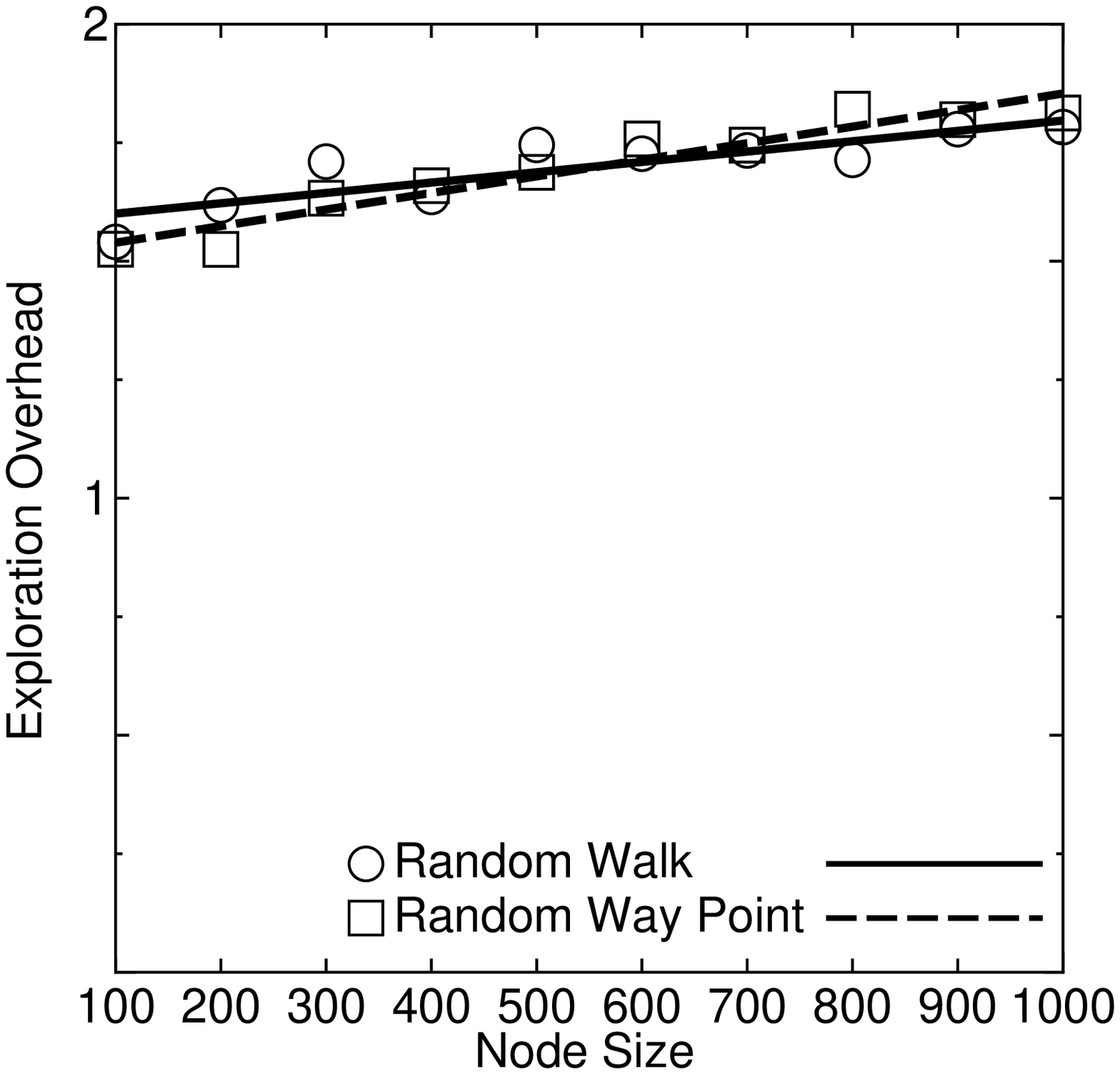}} \label{fig:3a}} \quad
      \subfigure[Impact of node speed] {\scalebox{0.42}{\includegraphics[width=\textwidth]{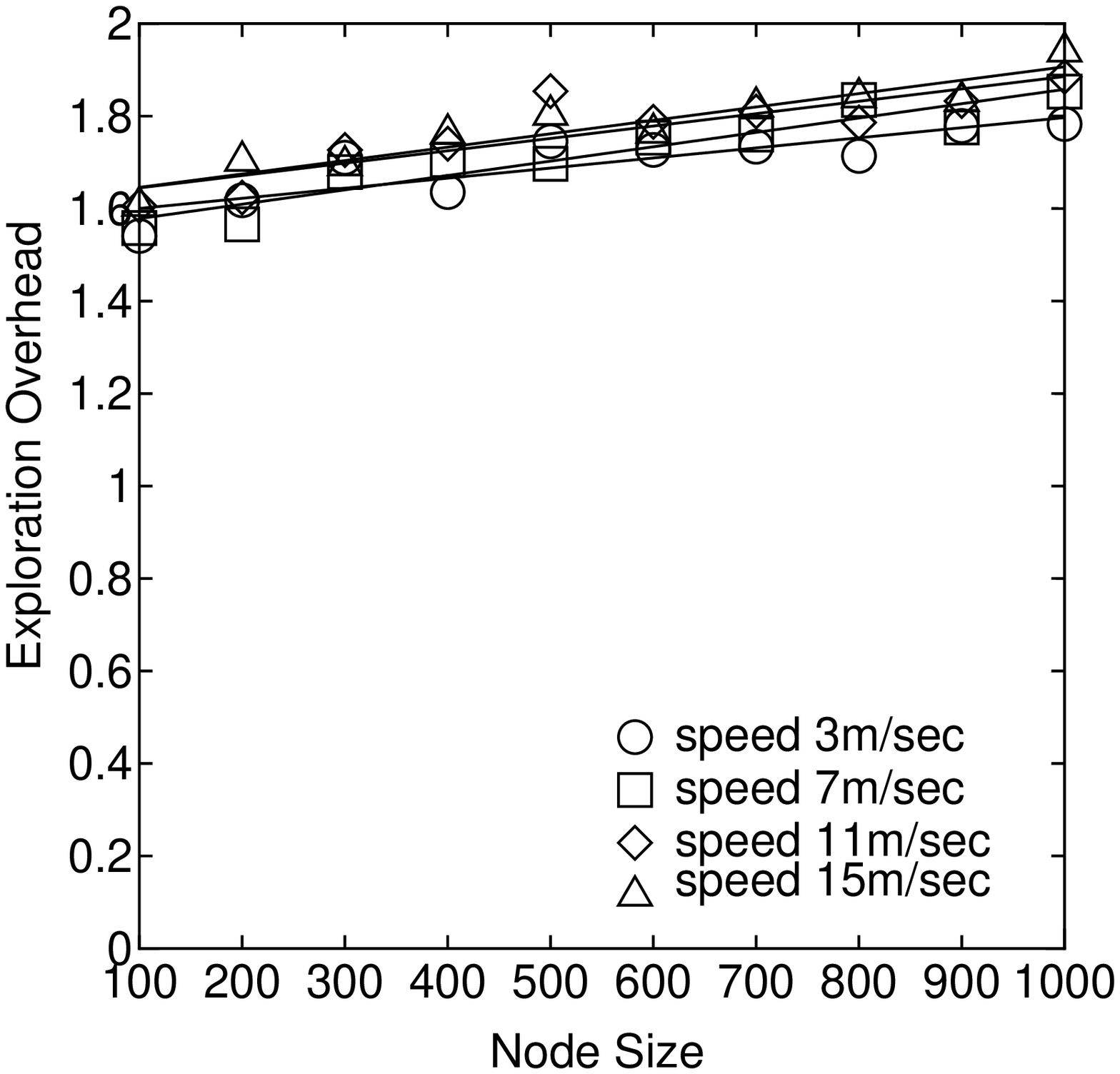}} \label{fig:3b}}          
      } 
    \vspace{-1mm}  
    \caption{Impact of mobility models and speed on a self-repelling random walk}
       \label{fig:3}
  \end{center}
\vspace*{-4mm}
\end{figure*}

\subsection{Exploration overhead}
In Fig.~\ref{fig:2a}, we plot the exploration overhead as a function of coverage percentage and observe that this ratio remains close to $1$ until about $80\%$ and then rises towards $2$. Further, in Fig.~\ref{fig:2b}, we see that the exploration overhead at $80\%$ coverage stays close to $1$ irrespective of network size. These results show that self-repelling random walks can achieve partial coverage in only $O(N)$ steps where $N$ is the number of nodes in the network. At $100\%$ coverage, we observe a small rise in exploration overhead as a function of network size. This trend is logarithmic indicating that the total number of steps to finish visiting all nodes grows as $O(Nlog(N))$. The exploration overhead stays under $2$ even at a network size of $1000$ nodes. These results show that self-repelling random walks can be used to traverse all nodes of a mobile ad-hoc network quite efficiently and consequently they can be used for applications such as data aggregation.

\subsection{Comparison with pure random walks}
In this subsection, we compare the performance of self-repelling random walks with pure random walks in which at each step, the next node to visit is chosen uniformly at random without consideration of the prior visits to the same node. In Fig.~\ref{fig:4a}, we plot the number of visits to each node until all nodes are visited at least once for a $500$ node network. In comparison with Fig.~\ref{fig:4b}, we observe that the tail of the distribution is much longer and the number of duplicate visits is much higher for pure random walks. Consequently, in Fig.~\ref{fig:5a}, we observe that the exploration overhead is significantly lower for self-repelling random walks as compared to pure random walks. In Fig.~\ref{fig:5b}, we see that even at partial coverage, pure random walks have a steadily rising exploration overhead that is much larger than $1$. On the other hand the exploration overhead for self-repelling random walks stays close to $1$ throughout.

\subsection{Impact of speed and mobility models}
In Fig.~\ref{fig:3a}, we show the exploration overhead at $100\%$ coverage for two different mobility models and observe that they are quite similar although the random waypoint mobility model does not maintain the uniformity in distribution of nodes in the MANET at all times. In Fig.~\ref{fig:3b}, we show the exploration overhead as a function of average node speed. We see that under high mobility, overhead is largely unaffected.

\section{Related Work}
\label{sec:related}
Random walks and their cover times (time taken to visit all nodes) have been studied extensively for different types of static graphs \cite{rw1, rw7}. Self-avoiding and self-repelling random walks are variants of random walks which bias the walk towards unvisited nodes. The unformity in coverage of such random walks in 2-d lattices has been pointed out in \cite{rw12}. Our paper extends the analysis of self-repelling random walks for application in MANETs that are modeled as time varying random geometric graphs. The idea of locally biasing random walks and its impact in speeding up coverage has been pointed out in \cite{rw8} for {\em static} networks. In contrast, this paper experimentally analyzes self-repelling random walks on top of mobile networks.

\section{Conclusions and future work}
\label{sec:discussion}
We have analyzed the coverage characteristics of self-repelling random walks in mobile ad-hoc networks under different mobility models and network sizes. Our results highlight the efficiency with which self-repelling random walks can be used to visit all nodes in a network and hence can be used for problems such as data aggregation without the need for expensive network structures. Our results have been verified on $2$ different mobility models and networks  upto $1000$ nodes at different node speeds. Until $85\%$ coverage, the exploration overhead remains close to $1$ and thus duplicate node visits are extremely rare. Even at $100\%$ coverage, the exploration overhead remains under $2$ for network sizes up to $1000$. 

One of our observations is that going from $85\%$ to $100\%$ coverage, the overhead rises somewhat steeply and the overhead also grows slowly with network size. We are exploring complementary strategies which can avoid this long tail in convergence and keep the exploration overhead close to $1$ irrespective of network size, thus resulting in an $O(n)$ time / messages to visit all nodes where $n$ is the number of nodes in the network.

\vspace{-3mm}
\bibliographystyle{unsrt}
\bibliography{vinod}


\end{document}